\documentclass[12pt]{article}
\begin{document}
\centerline{\bf {\Large Were the ancient Greeks right that   }}
\centerline{\bf {\Large  space is continuous material plenum?}}

\bigskip
\noindent I.E. Bulyzhenkov

\noindent {\it Lebedev Physical Institute, Leninsky pros. 53, Moscow 119991, Russia }

{\large
\begin{abstract}
All visible bodies are bound dense vertices of overlapping astroparticles with extremely weak $r^{-4}$ radial densities of elementary (and summary) matter beyond human perception and instrumental resolutions. The non-empty material space of the ancient Greeks have mathematical grounds in the self-consistent reading of Maxwell's phenomenology and Einstein's gravitation through continuous radial sources of classical fields. 
\end{abstract}

\bigskip
In trying to separate genetics from unknown information-reading me\-cha\-nisms, my brother called me one day to ask how physics can explain that his dog (ridgeback) repeatedly attacks TV images of lions (and only lions). How might this ridgeback from the Geneva club know that its ancestors were selected to fight with African lions? At that time I had no idea how to interpret space-time nonlocality of macroscopic matter in terms of classical fields and, therefore, replied with the on-duty quantum nonlocality of elementary particles, in the very terms one can find in `The Field' \cite {Lyn}. Physicists tend to replace all unexplained mysteries of macroscopic phenomena by sophisticated fluctuations of quantum particles. Conventional interpretations and the Standard Model are our self-defense against unexplained (and rarely published in peer-refereed journals) challenges of Nature. Nowadays, Kant might not get a chance to share with other thinkers that the mind has to possess {\it the synthetic a priori} or innate knowledge of `things in themselves' \cite {Kan} because a modern referee would like  to see all basic statements within the established scientific paradigm, rather than outside it. In this note I try to propose a logical bridge from the classical theory of electromagnetic and gravitational fields to non-empty material space of the ancient Greeks and cosmic life ideas of contemporary idealists.

Unexplained influences between distant events were already disputed by more than 65 world nations \cite {Lyn}. Here my personal experience is related mainly to private communications with Russian people, including Moscow scholars. Unlike the natural sciences, philosophy for a long time  was not considered by me as a powerful tool for the exploration of Nature. 
My teachers of physics avoided philosophy battles and non-rigorous definitions  like `matter is an objective reality given to us in perception' (the latter varies over humans and other live species). Instead, motivated by the tremendous triumph of the quantum formalism for probabilities of the delta-function particle and `observed' local collisions of accelerated charges, many contemporary physicists hardly accept holism for all matter in the Universe and distant correlations of nonlocal human beings. 

Contrary to leading modern scholars, multi-national population of the `materialistic' USSR, for example, always believed in intuition, distant communications of native souls, and a cosmic nature of marriages. 
Gurwisch's biofields \cite{Gur}, Vernadsky's Noosphere \cite {Ver} , Chizhevsky's pulsations of the Universe, Vasiliev's parapsychology, Kozyrev's astroenergy mirrors, Messing's psychic performances, Bekhterev's immortality of thoughts, and Russian cosmism philosophy \cite {Ser} were neither properly commented nor satisfactorily criticized   by the Russian Academy of Sciences. Instead of this, top scholars conventionally maintain point masses in warped space despite electric charges require only Euclidean space. Do masses and charges exist in different spaces of the same Universe? Or is there a common sense in such a point particle doctrine? Where could point-matter materialists fix cutting edges between observable inertial matter (an apple `given to us in perception') and its non-observable (beyond perception) parts, logically assigned to all observable bodies in the Universe by Kant-Hegel-Mach idealism? The substance surface cannot pass continuously through postulated empty space between postulated point material particles on the visual apple's edge. Once there is no matter in empty space beyond each atom, then there is no material line between two distant atoms, and there is no smooth material surface over the visible apple volume. 

It is unlikely that all idealists had no logic. Thus, why and how did they arrive at their `weird' conclusions about invisible realities of our physical world? Kant developed a deeper version of the celebrated Platonic forms (employed also by many materialists). Continuous distributions of invisible energetic substances over the ambient space were employed in Chinese and Hindu philosophies well before Plato's {\it Dialogues}, regarding eternal forms `residing in material objects'. Deductive logic, invented by Aristotle for a causal interpretation of Nature, had accepted that visual things have a lesser reality than Platonic forms. To Aristotle, substance is a combination of visible matter and some invisible forms, which are not a separate realm, nonetheless. He logically rejected the empty-space concept in favor of continuum space ({\it plenum}) filled everywhere with a background of invisible things.

Starting from Plato's {\it Dialogues}, the Western classical idealism culminated in Hegel's dialectic method for Nature with the absolute self-knowledge and collective mind-spirit-soul, denoted by one German word {\it geist} \cite {Heg}. Collective consciousness was coherently assumed by geochemist Vernadsky as the `sphere of human thought' (later called the {\it Noosphere}), which is the third phase of the Earth's formation after the Geosphere and the Biosphere \cite {Ver}. Is it possible to understand what part of the Standard Model a modern astrophysicist could specify behind Hegel's {\it geist}? May de Broglie waves, zero point fluctuations or strings be mentioned for Vernadsky's Noosphere? 

Physicists may surely talk about the invisible microworld and whatever they like, but philosophers discussed  macro and mega worlds where idealists insist on the global overlap of all minds in some information space with `the synthetic a priori knowledge'. The mind is not separated from matter in such an approach to space-plenum but can interact with matter through `transmutations of elements'. In other words, the material mind and other matter (including humans' bodies) should coexist in the same interaction space with a spatial overlap between the collective mind and the collective biomass of human beings. 

At first glance, this global overlap of everything seems to be absolute nonsense, as everyone sincerely testifies that mutual penetrations of live or inert matter apparently contradict to our daily observations, say separated localization of different apples on a table. Does this mean that German idealists had neither logic nor practical experience? Or that they (together with Mach, who was an expert in ballistics) did not understand classical mechanics for `localized' apples? Furthermore, genius Newton also maintained the `absurd' ether idea for the material overlap of gravitating bodies, Faraday considered charged field-matter around a charge's `center of force', Maxwell initiated the displacement current in terms of continuously charged flows of matter, and Clifford speculated on inhomogeneous material space. In the last century Mie, Hilbert, Einstein, Born, and Schwinger prompted a logical need in never observed continuous sources of classical fields. Have all these outstanding scholars lost logic for a moment? And have their opponents, like dialectic materialists, ever seen the amazing cohesiveness of bird flocks or fish schools when thousands of animals simultaneously change directions in timeless readings of what is going on? 

Now a reasonable question arises - who misinterpreted physical laws, deductive logic, or practical observations of matter in space? Quantum physics was irrelevant to the logical discovery of the ancient Greeks that observations of (macroscopic) bodies have a lesser reality than their eternal forms that are beyond the level of human perception. To the Greeks, the real space-plenum, hidden from detailed observation, is filled by continuous matter with nonlocal bounds, which were re-examined in the Einstein-Podolsky-Rosen paradox \cite {Ein}. Why not drop forever the unnecessary empty-space paradigm and the delta-operator source in Maxwell's and Einstein's field equations in favor of the ancient Greeks' continuous substance within the `perception fog' of non-empty space? Why not acknowledge Chinese yin-yang duality of material flows of particle-field densities or the conclusion of ancient Hindu philosophers that a body (its visual frames, to be precise) is an illusion of our incomplete observations of extended cosmic matter? All of these qualitative statements can be quantitatively replicated by Maxwell's Electrodynamics (ME) and Einstein's General Relativity (GR) when an analytical density of the radial particle replaces the Dirac delta density of the point particle. 

Einstein's 1915 equation, for example, states that the elementary source of Newtonian gravitational fields, $f \propto 1/r^{2} $, is the distributed energy density, $f^2 \propto 1 / r^{4} $, rather than a point mass. From here on, each Newtonian body, or carrier of mass-energy in modern gravitation equations, should be considered as the $r^{-4}$ radial astrodistribution of matter in the undivided and nonlocal Universe. Therefore, Einstein's equation with radial sources directly addresses Newton's overlapping bodies, Aristotle's space-plenum with invisible continuous things, and continuous energy-matter flows in  philosophies of the ancient East. 
Restricted frames of observed macroscopic bodies are perception illusions formed by concentrated bunches of dense vertices of $r^{-4}$ radial atoms (extending over the entire Universe). 

It is clear from the concept of nonlocal radial particles that when dialectic materialists cut vertices of continuous cosmic bodies above human perception clouds they unavoidably came to the opposite conclusion of Hegel, the father of the universal dialectic method. To materialists, ultimate reality is only visible matter above the perception level, while radial (invisible but still material) astrotails or fine formations of overlapping matter do not fall under their pragmatic definition of realities. If elementary particles were indeed localized point peculiarities in empty physical space rather than in rough mathematical models (with unphysical infinite energies), then the "popular" definition of matter through finite human perception might make some sense. But non-empty space distributions of matter below restricted human perception exist (due to their influence on observed events) and they are even more fundamental for Nature than visible bodies. For example, invisible material thoughts in the Noosphere and Kant-Hegel's Universe with the absolute self-knowledge of Nature continue to maintain steady energy-information forms (a dog repeatedly attacks TV images of never seen lions) even after the full disintegration of macroscopic bodies of former thinkers into gases of atoms.

Unfortunately for idealists, modern astrophysicists seem believe that once telescopes register the Moon "there" then there are no Moon's matter densities in marine tidal waves "here". Such a superficial approach to observations tends to  to simplify description of matter through unphysical delta-operators for presumably point particles within presumably empty space. However, this simplification contradicts the Hilbert-Einstein density of the (distributed) gravitational source in the Lagrange variational formalism for one carrier of matter.  In fact, the global overlap of nonlocal mass-energies of inert material objects and human beings seems contradictory to daily observations only due to a finite perception level. This `weird' overlap of all matter-energy, logically inferred by idealists, does exist in non-empty space for Einstein's gravitation and Maxwell's electrodynamics, where the nonlocal carrier of the elementary charge $e$ and mass $m$ can be analytically described \cite {Buly} by the continuous radial density $n(r) = r_o/4\pi r^2(r + r_o)^2$ instead of the delta operator $\delta (r)$ from the controversial model of localized particles. 
The global spatial overlap of all radial sources within a joint material space can provide mathematical grounds for  the ancient Greeks' logic and Kant's interpretation of Platonic forms. Euclidean space geometry, the Kant example of {\it a priori} knowledge in Nature, does not fail in Einstein-Machian astrophysics with overlapping radial sources \cite {Buly}. 

Mach \cite{Mac} indeed had analytical grounds for astrostates of inertial matter
and for criticism of localized Boltzman atoms because the true nature of `microscopic' atoms in Einstein's gravitation corresponds to the $r^{-4}$ radial sources distributed over the infinite Universe. Therefore, the relativistic mechanics of such continuous classical mass-energies should depend on their spatial overlap with all `distant' radial stars or with the Machian `rest of the Universe'. At the same time, the half-mass radius $r_o = Gm/c^2 \approx 10^{-57}m$    of the electron, for example, is much less than the top limit of space measurements $10^{-18} m$. Therefore the continuous electron was technically correctly measured in practice as a point mass-energy.   

The $r^{-4}$ nonlocal classical matter is rigorously defined through self-consis\-tent source-field solutions in Maxwell's and Einstein's equations, rather than through unspecified human perception. One fairly ought to admit that ancient philosophers interpreted real continuous bodies and cosmic nonlocality of human beings much better than contemporary physicists, equiped  by the international supercollider, national research labs and the theory of ME / GR fields (where continuous sources were erroneously replaced with point peculiarities due to misunderstanding of the nature of measurements in material space). The point is that all classical field equations can self-consistently accept the elementary radial particle integrated into the spatial structure of its field. The Mie/Einstein directive toward extended sources of  ME/GR fields may result in an advanced mathematical description of the global overlap of all inert and living bodies that was inferred by philosophers well before the formal, 'engineering' reading of model field equations through unreasonable postulated point sources. Continuous sources in Maxwell's physics and Mach-Einstein's relativism can analytically justify nonlocality of classical radial particles under causal relativistic motion of their geometrical vertices or Faraday's centers of force. 

Nowadays support of better science, if it is against the main financial stream, is not a common practice. I hardly remember who trully follows around the Aristotle's motto 'Amicus Plato, sed magis amica veritas'. Researchers back the idea that further measurements of space beyond the achieved limit of $10^{-18} m$ will facilitate a solution of the mass origin problem. But why we are forgetting to add that the mass creating chiral symmetry is violated at much larger scales, at least at $10^{-15}m$? The ancient Greeks would logically extended this symmetry violation on all other scales of their material space? Why not to grant a discussion floor to science critics of the empty-space paradigm rather than to serve silently to the delta-operator "innovation" for the continuous (in reality) electron?

Nonlocal correlations of bio-cells \cite{Kaz} and photons \cite{Asp} have already been proven in modern laboratories by the scientifically recognized methods. The extended mind and Noosphere collective consciousness are currently under successful investigation in many research programs based on web-coordinated experiments. An encountered number of facts has already confirmed the cosmic nature of living matter in our nonlocal world where everyone is a citizen of the entire Universe, rather than the small planet Earth. I strongly suggest replacing of the model, operator charge current $ {{\bf j} ({\bf r}) } = \sum_k e_k \delta({\bf r} - {\bf r}_k) {\dot {\bf r}}_k $ in Maxwell's equations with the analytical continuous current ${{\bf j} ({\bf r}) } = \sum_k e_k n(|{\bf r}- {\bf r}_k |){\dot {\bf r}}_k $ of  nonlocal radial charges in any field point ${\bf r}$. I encourage empty space materialists to revise their point source approximations of elementary radial matter in favor of analytical descriptions of the continuous classical charge, distant mind-matter correlations in the  collective cosmic life, and nonlocal phenomena of radial particles in non-empty material space of the ancient Greeks, the German idealists \cite{Kan,Heg,Mac}  and the Russian cosmism  philosophers \cite{Ver,Ser,Kaz}.

} 

\newpage

\end {document}